\documentclass[aps,preprint,a4paper,showpacs,showkeys,superscriptaddress]{revtex4-1}
\ifx\pdfoutput\undefined
\usepackage[usenames]{color} 
\usepackage[dvips]{graphicx}
\else
\usepackage{color} 
\definecolor{BrickRed}{rgb}{0.85,0.15,0.25}
\definecolor{MidnightBlue}{rgb}{0,0.45,0.85}
\definecolor{ForestGreen}{rgb}{0,0.85,0.45}
\usepackage{graphicx}
\usepackage{epstopdf}
\fi
\usepackage{subfigure}
\usepackage{latexsym,amsmath,amssymb}
\usepackage{hyperref}     
\hypersetup{colorlinks,%
  citecolor=blue,%
  linkcolor=cyan,%
  pdftex}
\allowdisplaybreaks

\usepackage{acronym}

\usepackage{umoline}
\newsavebox\CBox

\begin{document}


\title{Origin of Hawking radiation: firewall or atmosphere?}



\author{Wontae Kim}%
\email[]{wtkim@sogang.ac.kr}%
\affiliation{Department of Physics, Sogang University,
Seoul 04107, South Korea}%

\date{\today}

\begin{abstract}
The Unruh vacuum not admitting any outgoing flux at the horizon implies that
the origin of the outgoing Hawking radiation is the atmosphere of a near-horizon quantum
region without resort to the firewall; however,
the existence of the firewall of superplanckian excitations at the horizon
can be supported by the infinite Tolman temperature at the horizon.
In an exactly soluble model,
we explicitly show that the firewall necessarily emerges out of the Unruh vacuum
so that the Tolman temperature in the Unruh vacuum is divergent
in essence due to the infinitely blueshifted negative ingoing flux crossing the horizon
rather than the outgoing flux.
We also show that
the outgoing Hawking radiation in the Unruh vacuum indeed originates from the atmosphere,
not just at the horizon, which is of no relevance to the infinite blueshift.
Consequently, the firewall from the infinite Tolman temperature
and the Hawking radiation from the atmosphere turn out to be compatible,
once we waive the claim that the Hawking radiation in the Unruh vacuum originates from the infinitely blueshifted
outgoing excitations at the horizon.

\end{abstract}



\maketitle


\acrodef{AMPS}[AMPS]{Almheiri, Marolf, Polchinski and Sully}
\acrodef{CGHS}[CGHS]{Callan-Giddings-Harvey-Strominger}
\acrodef{RST}[RST]{Russo-Susskind-Thorlacius}

\section{Introduction}
\label{sec:intro}
Hawking radiation as an information carrier \cite{Hawking:1974sw} is subject to observers,
which requires black hole complementarity such that
there are no contradictory physical observations between different observers
\cite{Susskind:1993if}.
Recently, the firewall paradox~\cite{Almheiri:2012rt} (or
the energetic curtain from some different assumptions
\cite{Braunstein:2009my})
was that the assumptions of the purity of the Hawking radiation, the semiclassical quantum field theory,
and the equivalence principle are not consistent with one another.
Then they suggested a conservative solution to this paradox that the infalling observer when crossing the horizon
could find the firewall of high frequency quanta beyond the Planckian scale
after the Page time \cite{Page:1993wv}.
Subsequently, there have been many
arguments for this paradox and alternatives \cite{Bousso:2012as, Nomura:2012sw, Susskind:2012rm, Hossenfelder:2012mr, Giddings:2013kcj,
 Almheiri:2013hfa, Hutchinson:2013kka, Freivogel:2014dca}.

On the other hand, Unruh
attained a startling conclusion that Hawking radiation
appears in the absence of the outgoing flux at the horizon
\cite{Unruh:1976db} and also
showed that the process of thermal particle creation is low energy behavior
and the highest frequency mode does not matter for the thermal emission by
using a modification of the dispersion relation in a sonic black hole numerically \cite{Unruh:1994zw}.
Moreover, it was shown that the
effective blueshift of the outgoing Hawking radiation
remains finite
by averaging out the Tolman factor outside the horizon in terms of the Poisson distribution
\cite{Casadio:2002dj}.
All these imply that the Hawking radiation originates from a macroscopic distance
outside the horizon.
Interestingly, it was also claimed that the Hawking radiation can be retrieved by an alternative scenario
that a positive outward flux at the horizon is interpreted as the negative influx
without recourse to a pair creation scenario \cite{Israel:2015ava}.

Recently,
there was a refined question concerning the origin of the Hawking radiation in the Unruh vacuum
by Giddings \cite{Giddings:2015uzr},
where the evidence concerns the three relevant points that are: (1) the effective emitting area of Hawking radiation is
considerably larger than the size defined by the area of the black hole~\cite{Page:1976df},
(2) there is no outgoing flux at the horizon in the Unruh vacuum
and there should appear a transition from the ingoing to the outgoing flux
over a large quantum region, and
(3) the size of wave length of a thermal Hawking particle is larger than
the horizon size.
One of the essential ingredients is that
the transition from the ingoing to the outgoing flux should appear over the atmosphere of
the quantum region outside the horizon
without resort to the firewall.
Moreover, the importance of
the atmosphere
was also emphasized in connection with the nonviolent scenarios
for the information loss paradox~\cite{Giddings:2006sj}.

However, the existence of the firewall seems obvious
from the fact that the Tolman temperature defined in the Unruh vacuum
is infinite on the horizon \cite{Tolman:1930zza},
which is indeed due to the infinite blueshift of the Hawking temperature there.
In other words, the thermal Hawking particles at infinity
might be ascribed to the infinitely blueshifted outgoing radiation at the horizon or very
near the horizon.
This argument, as mentioned above, would not be reliable, since the Unruh vacuum
does not admit any outgoing flux on the horizon semiclassically \cite{Unruh:1976db}.
So it is not likely to get the firewall from the outgoing flux, and thus
it might be tempting to conclude
that the firewall is incompatible with the Unruh vacuum.
In these regards, the origin of Hawking radiation and the reason for the
existence of the firewall as well as their relationship
still seem to be equivocal in spite of many efforts.

In this work, we elucidate how the Hawking radiation and the firewall appear
simultaneously in a tractable field theoretic model,
and then provide a compelling argument for their compatibility between
the firewall and the Hawking radiation.
The key is to decompose quantitatively the Tolman temperature
read off from the Stefan-Boltzmann law
into the two chiral temperatures of $T_\text{L}$, $T_\text{R}$
defined by the negative influx and the positive outward flux, respectively, and then identify
their properties carefully.
It will be shown that $T_\text{L}$ becomes infinite at the horizon, which is regarded as a signal of the
firewall; however, it vanishes at infinity, so that it does not affect the asymptotic observer at infinity.
The essential reason for the existence of the firewall is
due to the infinitely blueshifted negative influx crossing the horizon
rather than the outward flux.
On the other hand, $T_\text{R}$ will be shown to be finite everywhere by identifying it
with a newly derived effective Tolman temperature from the modified Stefan-Boltzmann law.
In particular, it vanishes at the horizon and approaches the Hawking temperature
at infinity. Thus, it shows that the
outgoing Hawking radiation originates from the atmosphere of the near-horizon quantum region,
not just at the horizon.
After all, the present analysis in the semiclassically fixed background approximation will show
that the firewall is not only a conservative but also a natural solution in the Unruh vacuum,
and the Hawking radiation indeed originates from the atmosphere without any conflicts with the firewall.

The organization of this paper is as follows.
In Sec. \ref{sec:Energy density and flux},
we introduce the quantized stress tensor for a single scalar field
on a general class of two-dimensional black hole background without choosing a specific metric,
and discuss black-hole vacua.
In Sec. \ref{sec:Stefan-Boltzmann law in equilibrium},
in the Israel-Hartle-Hawking vacuum \cite{Hartle:1976tp,Israel:1976ur},
we recapitulate the recent formulation of the effective Tolman temperature \cite{Gim:2015era}
and then write it in terms of a more preferable form in order to serve our purpose.
In thermal equilibrium, the effective Tolman temperature turns out to be a natural generalization
of the usual Tolman temperature in the presence of
conformal anomaly \cite{Deser:1976yx} responsible for the Hawking radiation \cite{Christensen:1977jc}.
In Sec. \ref{sb}, the
Tolman temperature defined in the Unruh vacuum is decomposed into the left and right temperatures
from the Stefan-Boltzmann law for a radiating system by using the results of the preceding sections.
Then the arguments for the firewall and the atmosphere
are explicitly discussed by choosing a specific
metric. Finally, conclusion and discussion are given in Sec. \ref{sec:Diss}.

\section{Energy density and flux}
\label{sec:Energy density and flux}

Let us start with a two-dimensional general static black hole described by the metric,
\begin{eqnarray}
ds^2=- g(r)dt^2+\frac{1}{g(r)}dr^2,
\end{eqnarray}
where $g(r)$ is an asymptotically flat metric function. The constants are set to $\hbar=k_{\text B}= G=c=1$.
The event horizon $r_\text{H}$ is defined by $g(r_\text{H})$=0, and
the Hawking temperature is calculated from the definition of the surface gravity as
$T_\text{H}=g'(r_\text{H})/4\pi$  \cite{Hawking:1974sw}
where the prime denotes the derivative with respect to $r$.
The Hawking temperature is blueshifted
for a distant observer outside the horizon \cite{Wald:1999xu}, which is simply
written as the Tolman form \cite{Tolman:1930zza}.

From the covariant conservation law and the conformal anomaly of $\langle T^\mu_\mu \rangle= R/(24\pi)$
for a two-dimensional massless scalar field
\cite{Deser:1976yx}, the components of the stress tensor are determined as
$\langle T_{\pm\pm} \rangle =\left(gg''-({1}/{2})g'^2 +t_\pm  \right)/{(96\pi)}$,
$\langle T_{+-} \rangle =gg''/{(96\pi)}$,
where $t_\pm$ reflect the non-locality of the conformal anomaly \cite{Christensen:1977jc}.
Note that the expectation value of the energy-momentum tensor was
written by using the tortoise coordinates written in the form of light-cone
where $\sigma^{\pm}=t \pm r^{*}(r)$ and $r^*=\int dr/g(r)$,
but $g(r)$ was just written in terms of $r$ instead of $r^*$ for convenience \cite{Giddings:2015uzr}.
Then, the proper energy density, pressure, and flux can be defined as
$\varepsilon=\langle T_{\mu\nu}\rangle u^\mu u^\nu $, $p=\langle T_{\mu \nu} \rangle n^\mu n^\nu$, and ${\cal{F}}=-\langle T_{\mu\nu} \rangle u^\mu n^\nu$,
where $u^\mu$ is a two-velocity and $n^\mu$ is a spacelike unit normal vector satisfying
$n^\mu n_\mu=1$ and $n^\mu u_\mu=0$.
Explicitly, in the light-cone coordinates,
the velocity vector from the geodesic equation of motion
and the normal vector are solved in a freely falling frame from rest as \cite{Eune:2014eka}
\begin{equation}
u^+ =u^- = n^+ =- n^- =\frac{1}{\sqrt{g}}. \label{n}
\end{equation}
In particular,
the proper energy density and flux are expressed by
\begin{eqnarray}
\varepsilon &=& \frac{1}{g} \left( \langle T_{++} \rangle +\langle T_{--} \rangle + 2 \langle T_{+-} \rangle \right) \label{rho}, \\
{\cal{F}} &=& -\frac{1}{g} \left( \langle T_{++} \rangle -\langle T_{--} \rangle  \right), \label{flux}
\end{eqnarray}
where the redundant pressure is related to the energy density via
the trace relation of $\langle T^\mu_\mu \rangle=-\varepsilon + p$.
From Eqs. \eqref{rho} and \eqref{flux}, the energy density and flux are explicitly written
as $\varepsilon = \left(4gg''-g'^2 +t_+ +t_- \right)/(96\pi g)$ and
${\cal{F}}  = - \left( t_+ - t_- \right)/(96\pi g)$.

In the Israel-Hartle-Hawking vacuum \cite{Hartle:1976tp,Israel:1976ur},
the stress tensor is regular at both the future horizon and the past horizon,
so that the regularity condition determines the
integration constants as $t_+=t_-=(1/2)g'^2(r_\text{H})$.
Let us assume that the metric function is finite at least up to the second derivative with $g'' <0$, for
instance, which holds for the Schwarzschild black hole or the CGHS black hole \cite{Callan:1992rs},
then the curvature scalar of $R=-g''$ is positive finite.
The proper energy density \eqref{rho} also becomes finite everywhere. In particular,
it is negative finite at the horizon, $\varepsilon_\text{HH}(r_\text{H})=g''(r_\text{H})/(48\pi)$,
while it is positive finite at
infinity, $\varepsilon_\text{HH}(\infty)= g'^2(r_\text{H})/(96\pi)$. It shows that the proper energy density is not always positive.

There is another equilibrium state defined by $t_+=t_-=0$
called the Boulware vacuum \cite{Boulware:1974dm}.
The energy density \eqref{rho} is negatively divergent at the horizon, $\varepsilon_{\text B}(r_\text{H}) \to -\infty$
and negatively vanishes at infinity, $\varepsilon_{\text B}(\infty) =0$.
If such a black hole exists, then it will be surrounded by the negative energy density in equilibrium.
Note that the energy density is divergent at the horizon,
so that the smoothness of the horizon is not warranted.

The Unruh vacuum of our interest is considered
by twisting two equilibrium states asymmetrically
in such a way that $t_+ =0$ and $t_- =g'^2(r_\text{H})/2$ in order to describe an evaporating black hole
semiclassically \cite{Unruh:1976db}.
Then the proper energy density is negative infinity
at the horizon, $\varepsilon_{\text U}(r_\text{H}) \to -\infty$ like the case of the Boulware vacuum defined by $t_+=t_-=0$
\cite{Boulware:1974dm}.
It is positive finite at infinity like the case of the Israel-Hartle-Hawking vacuum
defined by $t_+=t_-=(1/2)g'^2(r_\text{H})$ \cite{Hartle:1976tp,Israel:1976ur}, but its magnitude is half
of that of the Israel-Hartle-Hawking vacuum, i.e., $\varepsilon_\text{U}(\infty)= g'^2(r_\text{H})/(192\pi)=\varepsilon_\text{HH}(\infty)/2$.
As expected, the non-vanishing flux is obtained as ${\cal{F}}_\text{U}(r)= g'^2(r)/(192\pi g)$ which
is coincident with the energy density at infinity, but it is
divergent at the horizon.

Note that the above expectation value
of the energy-momentum tensor $\langle T_{\mu\nu} \rangle$  in any vacua is finite at the horizon,
whereas the proper flux and the energy density in the Unruh vacuum are divergent there.
One might wonder what the origin of the divergence is.
The expectation value of the energy-momentum tensor
$\langle T_{ab}(\xi^a,\xi^b) \rangle$ defined in a locally inertial coordinate system
can be obtained from the general coordinate transformation of
$ \langle T_{\mu\nu} \rangle$ defined in the tortoise coordinate system of $\sigma^{\pm}=t \pm r^{*}(r)$,
which is implemented by
\begin{equation}
\langle T_{ab} \rangle=\frac{\partial \sigma^\mu}{\partial \xi^a}
\frac{\partial \sigma^\nu}{\partial \xi^b} \langle T_{\mu\nu} \rangle,
\label{tensor}
\end{equation}
where $a,b=0,1$ are the indices for the locally inertial coordinate system and
$\mu,\nu=\pm$ are for the tortoise coordinate system.
We also consider the coordinate transformation of the velocity vector and the unit normal spacelike vector as
\begin{equation}
u^\mu = \frac{\partial \sigma^\mu}{ \partial \xi^a } u^a=\frac{\partial \sigma^\mu}{ \partial \xi^0 },~~~~~~
n^\mu = \frac{\partial \sigma^\mu}{ \partial \xi^a } n^a=\frac{\partial \sigma^\mu}{ \partial \xi^1 } \label{velocity}
\end{equation}
where $u^a=(1,0), n^a=(0,1)$ are defined in the locally inertial coordinate system.
Therefore, the flux ${\cal{F}}=-\langle T_{01} \rangle$  from Eq. \eqref{tensor} is written as
\begin{equation}
-\langle T_{01} \rangle=-\frac{\partial \sigma^\mu}{\partial \xi^0} \frac{\partial
\sigma^\nu}{\partial \xi^1} \langle T_{\mu\nu} \rangle=-u^\mu n^\nu
\langle T_{\mu\nu} \rangle, \label{t01}
\end{equation}
by using Eqs. \eqref{velocity}, which is nothing but Eq. \eqref{flux} when Eq.\eqref{n} is used.
The energy density \eqref{rho} can also be obtained from the coordinate transformation of
$\langle T_{00} \rangle=(\partial \sigma^\mu /\partial \xi^0) (\partial \sigma^\nu /\partial \xi^0)
\langle T_{\mu\nu} \rangle=u^\mu u^\nu
\langle T_{\mu\nu} \rangle$, which reproduces Eq. \eqref{rho}.
As a result, in the Unruh vacuum, the divergent proper flux
at the horizon is due to the singular coordinate transformation
at the horizon.


\section{Stefan-Boltzmann law in equilibrium}
\label{sec:Stefan-Boltzmann law in equilibrium}

Conventionally, the Stefan-Boltzmann law in thermal equilibrium rests upon
the traceless condition of the stress tensor~\cite{Tolman:1930zza};
however, it is worth noting that the trace anomaly is responsible for the
Hawking radiation~\cite{Christensen:1977jc}.
In that sense, assuming the nontrivial trace of the stress tensor,
one should obtain a modified Stefan-Boltzmann law which gives an effective Tolman temperature
induced by the trace anomaly \cite{Gim:2015era}.
Let us now obtain the effective Tolman temperature from the modified Stefan-Boltzmann law
in the Israel-Hartle-Hawking vacuum prior to the discussion in the case of the Unruh vacuum,
and then write
the effective temperature in terms of a much more convenient form for our purpose.

In a proper frame, the first law of thermodynamics reads as $dU =TdS-pdV$,
where $U$, $T$, $S$, and $V$ are thermodynamic internal energy,
temperature, entropy, and volume of a system in thermal equilibrium.
Inserting the internal energy defined as $U=\int \varepsilon dV$
and the Maxwell relation of
$(\partial S/\partial V )_T = (\partial p/\partial T)_V$
into the differentiated form of the thermodynamic first law with respect to the volume,
one can obtain
\begin{equation}
\label{key}
T \left(\frac{\partial \varepsilon}{\partial T}\right)_V-2\varepsilon = \langle T^\mu_\mu \rangle,
\end{equation}
where the trace anomaly is independent of the temperature
~\cite{BoschiFilho:1991xz}.
From Eq.~\eqref{key}, the energy density is immediately solved as
\begin{align}
 \varepsilon= \gamma T^2_\text{eff} -\frac{1}{2} \langle T^\mu_\mu \rangle  \label{effenergy}
\end{align}
where $T_{\rm eff}$ is the effective Tolman temperature.
The integration constant $\gamma$ in Eq. \eqref{key}
is the Stefan-Boltzmann constant given as $\gamma = \pi/6$
for a massless scalar field~\cite{Christensen:1977jc}.
The modified Stefan-Boltzmann law \eqref{effenergy}
simply reduces to the usual Stefan-Boltzmann law of $\varepsilon=\gamma T^2$
for the traceless case, which yields the usual Tolman temperature.

Note that
the proper energy density in the Israel-Hartle-Hawking vacuum is not always positive \cite{Visser:1996ix}.
Thus, one can find
the position $r_0$ where the energy density vanishes,
so that the region of the positive energy density is separated from that of the negative energy density.
For example, the position is estimated as $r_{0} \sim 2.98 M$
in the two-dimensional Schwarzschild black hole \cite{Eune:2014eka}.
It should be noted that
the usual Stefan-Boltzmann law holds only at infinity such as $\varepsilon_\text{HH}(\infty)=\gamma T_{\text H}^2$
; however, it does not appear to be
reliable extremely in
the region of the negative energy density for $r_{\text H} <r<r_0$.
Fortunately, the energy density need not
be positive thanks to the anomalous term in the modified Stefan-Boltzmann law \eqref{effenergy}.

Now, plugging the expression for the energy density in Eq.~\eqref{rho} into Eq.~\eqref{effenergy},
one can get the effective Tolman temperature in the Israel-Hartle-Hawking vacuum as
\begin{eqnarray}
\label{eff}
\gamma T^2_\text{eff} =\frac{1}{g} \left( \langle T_{++} \rangle_{\text{HH}} + \langle T_{--} \rangle_{\text{HH}} \right)
 = \frac{1}{96\pi g}\left( 2gg''-g'^2 +g'^2(r_\text{H}) \right),
\end{eqnarray}
where the stress tensors are calculated with respect to the Israel-Hartle-Hawking vacuum.
In contrast to the divergent usual Tolman temperature at the horizon,
the effective Tolman temperature \eqref{eff} vanishes at the horizon,
which can be shown by taking the limit at the horizon.
In fact, there is neither influx nor outward flux at the horizon, and; thus, there is no reason for the firewall to exist.
This fact is consistent with the regularity of the renormalized stress tensor at the horizon
in the Israel-Hartle-Hawking vacuum \cite{Visser:1996ix} and
the explicit calculation to make use of the detector \cite{Singleton:2011vh}.

\section{Stefan-Boltzmann law in a radiating system}
\label{sb}

Let us now derive the temperature for a radiating system such as the black hole in the Unruh vacuum.
The temperature is obtained from
the radiated power which is just the proper flux. The proper temperature
in the Unruh vacuum can be read off from
the two-dimensional Stefan-Boltzmann law given by Giddings \cite{Giddings:2015uzr},
\begin{equation}
\label{Giddings}
\sigma T^2 =-\frac{1}{g} \left(\langle T_{++} \rangle_{\text U} -\langle T_{--} \rangle_{\text U} \right)
=\frac{\pi}{12}\left(  \frac{T_\text{H} }{ \sqrt{g}} \right)^2,
\end{equation}
where the stress tensors are calculated with respect to the Unruh vacuum.
The two-dimensional Stefan-Boltzmann constant in a radiating system $\sigma$
is half of that of the equilibrium state, so that $\sigma=\gamma /2=\pi/12$ is consistent.
Note that the ingoing and outgoing stress tensors in the Unruh vacuum are
the same as those in the Boulware and Israel-Hartle-Hawking vacua, respectively.
So the black hole temperature in the Unruh vacuum \eqref{Giddings} can be decomposed into
the left and right chiral temperatures
without mixing chirality as
\begin{eqnarray}
\sigma T_\text{L}^2 =-\frac{\langle T_{++}\rangle_{\text{B}}   } {g},~~~~~~  \sigma T_\text{R}^2 =\frac{\langle T_{--} \rangle
_{\text{HH}}}{g}, \label{leftright}
\end{eqnarray}
where the stress tensors of $\langle T_{++}\rangle_{\text{B}}$ and $\langle T_{--} \rangle_{\text{HH}}$ are
defined with respect to the Boulware and Israel-Hartle-Hawking vacua, respectively.

Equilibrium states of black holes are commonly described by the  Boulware and Israel-Hartle-Hawking vacua.
This is possible only when the systems are locally in equilibrium and sufficiently slowly varying.
 In contrast to these vacua,
 the net flux is not zero for the Unruh vacuum,
  so that the black hole in this state is not in equilibrium as seen from Eq. \eqref{Giddings}.
   If the two equilibrium systems are interacting,
   then the thermal temperature for a better interpretation will require a quasi-equilibrium condition
    between the two different equilibrium systems. However, in the present semiclassical approximation,
    the influx and outward flux are actually decoupled and do not interfere with the influx and outward flux.

For the left temperature in Eq. \eqref{leftright}, the ingoing flux is negative finite at
the horizon as $\langle T_{++}(r_{\text H}) \rangle_{\text B}=-g'^2(r_\text{H}) /(192\pi)$,
so that the left temperature becomes positively divergent at the horizon as $T_\text{L} (r_H) \rightarrow +\infty$.
So the firewall in the Unruh vacuum arises from
the infinite blueshift of the negative ingoing flux despite the absence of the outgoing flux at the horizon,
so that the Tolman temperature is consequently divergent at the horizon.
However, this ingoing flux decreases to zero and does not
reach infinity, so that $T_\text{L}(\infty)=0$.
It implies that the ingoing superplanckian excitations have no impact on the
asymptotic observer at infinity. Thus, these excitations are certainly responsible for the firewall
but completely decoupled from the Hawking radiation at infinity.

Next, after rescaling $\sigma  =\gamma /2$, the right temperature in Eq. \eqref{leftright}
can be shown to be equivalent to the effective Tolman temperature \eqref{eff},
\begin{align}
T_\text{R}=T_\text{eff}, \label{good}
\end{align}
by use of the equilibrium condition of $\langle T_{++}\rangle_{\text{HH}}=\langle T_{--}\rangle_{\text{HH}} $.
Thus, the right temperature
directly possesses the same properties as those of the effective Tolman temperature,
so that it is finite everywhere.
In particular, it vanishes at the horizon and approaches the Hawking temperature exactly
at the asymptotic infinity.
Therefore, it shows that the Hawking particles at infinity
originate from the atmosphere outside the horizon
rather than the firewall.

On the other hand, from the left temperature in Eq. \eqref{leftright},
it is of interest to note that one can also define a black hole temperature
in the Boulware vacuum in a manner similar to the way of the Israel-Hartle-Hawking vacuum
by replacing $\sigma  = \gamma /2$ and
$\langle T_{++}\rangle_{\text{B}} =\left(\langle T_{++}\rangle_{\text{B}} +\langle T_{--}\rangle_{\text{B}}\right)/2 $
since $\langle T_{++}\rangle_{\text{B}}=\langle T_{--}\rangle_{\text{B}} $.
So, the left-right temperatures in Eq.~\eqref{leftright} can be compactly written in the unified manner as
\begin{align}
\label{wonder}
\gamma T^2_\text{L,R} =\mp \frac{1}{g} \left( \langle T_{++} \rangle_{\text{B,HH}} + \langle T_{--} \rangle_{\text{B,HH}} \right),
\end{align}
where the local Boulware temperature and the local Hawking temperature
are eventually on an equal footing, and thus, the ingoing particles are also in thermal states like the outgoing particles.
It implies that the particles in the same chirality can be entangled with their partners,
but the particles in a different chirality need not be entangled since they are not
created from pair creation \cite{Israel:2015ava}.
Importantly, this would be one of the advantages of the present analysis without recourse to pair creation,
which could respect the monogamy principle in quantum mechanics.
\begin{figure}[pt]
  \begin{center}
  \includegraphics[width=0.45\textwidth]{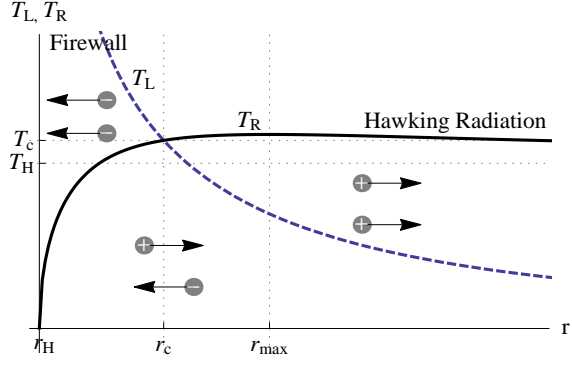}
  \end{center}
  \caption{The dashed curve is for $T_{\text L}$, and
   the solid one is for $T_{\text R}$, where $M=1$ for simplicity.
   The minus-plus signs in the small circles with the left and right arrows mean the negative influx
   and the positive outward flux, respectively.
      The critical position of the flux transition occurs at
    $r_{\text c} \sim 3.26M$ and the corresponding critical temperature $T_{\text c }$ is
   slightly higher than the value of the Hawking temperature $T_{\text H}$, where
   $T_c /T_{\text H} \sim 1.14$. The maximum of the right
   temperature occurs at $r_{\text{max}} \sim 4.32M$.
   }
  \label{fig:1}
\end{figure}

So far, the origin of the Hawking radiation and
the reason for the existence of the firewall have been discussed based on the generic metric.
Let us now discuss the arguments explicitly
in the two-dimensional Schwarzschild black hole
described by $g(r)=1- 2M/r$.
From Eq. \eqref{wonder}, one can obtain the left and right temperatures
as
\begin{align}
T_{\rm L}&=\frac{1}{\sqrt{2}\pi r} \sqrt{\frac{M}{r-2M}\left(1-\frac{3M}{2r}\right)}, \label{L}\\
T_{\rm R}&=\frac{1}{8\pi M} \sqrt{1+\frac{2M}{r}+\left(\frac{2M}{r}\right)^2-3\left(\frac{2M}{r}\right)^3}. \label{R}
\end{align}
Note that the left temperature is found to be infinite at the horizon
and vanishes at infinity, which means that the firewall appears at the horizon and
the asymptotic observer is free from the impact of
the firewall.
On the other hand, the right temperature is finite everywhere,
and it shows that the outgoing radiation is a very low energy phenomenon since it is almost comparable
to the Hawking temperature over the entire region outside the horizon.
In particular, it vanishes at the horizon and approaches the Hawking temperature at infinity.
As a corollary, from Eqs. \eqref{L} and \eqref{R} the Tolman temperature is obtained as $T= \sqrt{ T_\text{L}^2 + T_\text{R}^2}$
which shows only the
collective behavior of the two chiral temperatures.

In Fig. \ref{fig:1}, one may define a critical position $r_c$ at which
the two excitation energies are the same as $T_{\text L}=T_{\text R}$. It yields $r_{\text c} \sim 3.26M$ which
is a larger macroscopic distance as compared to the horizon size.
For $r <r_{\text c}$, the left temperature is dominant and
eventually predicts the firewall at the horizon, whereas
the right temperature is dominant for $r >r_{\text c}$ and reproduces the Hawking temperature at infinity.
Moreover, $T_{\text R}$ has a peak at $r_{\text{max}} \sim 4.32M$
which is larger than $r_{\text c}$. Thus, the critical transition from the influx to the outward flux occurs
before the right temperature arrives at the peak.

\section{Conclusion and discussion}
\label{sec:Diss}
In conclusion, a glimpse of the Tolman temperature in the Unruh vacuum
would lead to the conclusion that
Hawking radiation at infinity comes
from the infinitely blueshifted outgoing Hawking excitations at the horizon;
however, it was a misleading interpretation due to the two overlapped features.
In the context of the stress tensor calculations in this work,
this issue was clarified by decomposing the Tolman temperature in the Unruh vacuum
into the left and right chiral temperatures.
We showed that the firewall
in the Unruh vacuum comes from the negative influx crossing the horizon
rather than the outgoing flux at the horizon.
On the other hand, we also obtained that the Hawking radiation in the Unruh vacuum
comes from the positive
outward flux in the near-horizon quantum region of the atmosphere, not right at the horizon.
The right temperature became finite everywhere, so that
the low energy Hawking particles turned out to be irrelevant to the infinite blueshift.
Consequently,
the present study shows that the firewall from the infinite Tolman temperature at the horizon
and the Hawking radiation from the atmosphere outside the horizon are compatible,
if we discard the claim that the Hawking radiation in the Unruh vacuum
originates from the infinitely blueshifted
outgoing excitations at the horizon.

As for the information loss paradox, the present firewall need not play
a role of the entanglement-breaker between two partners of each pair across the horizon
created from pair production, but it is due to the infinite blueshift of
the influx at the horizon. So,
the firewall might be regularized in certain ways in order
to save the violation of the equivalence principle at the horizon.
Regarding the firewall,
it was claimed that there are no event horizons
\cite{Hawking:2014tga}, and they are inappropriate to describe the evaporating
black hole practically \cite{Visser:2014zqa}.
In addition, it was shown that the quantum back reaction of the geometry
renders a star stop collapsing a finite radius larger than
its horizon \cite{Mersini-Houghton:2014zka, Mersini-Houghton:2014cta}.
Recently, it was also claimed that the firewall is due to the limitation of
the semiclassically fixed background \cite{Nomura:2016qum}.
In fact, the present study was also performed on such a fixed
background with the event horizon.
In these regards, it could be speculated that the firewall in the Unruh vacuum might be
smoothed out by taking into account the quantum back reaction of the geometry
going beyond semiclassical approximations, which remains unsolved.

In connection with the above comment,
the authors \cite{Nomura:2016qum} argued that the firewall paradox manifests as a limitation of semiclassical treatment
in that the semiclassical theory possesses an unphysically large Fock space
built by creation and annihilation operators on a fixed black hole background.
The firewall is in essence the divergent quantity in the presence of the event horizon
on the fixed background. In these regards, the firewall should be regularized
by taking into account appropriate back reaction of geometry, and then
I just speculate that the ingoing modes might be a comparable wavelength to that of the outgoing flux.
However, the process of information recovery is another issue,
whether the nonlocal quantum field theory advocated by Giddings \cite{Giddings:2013kcj}  is essential or not;
otherwise, the other mechanisms responsible for the information retrieval might be needed.

The final comments are now in order.
Firstly, the decomposition of left-right fluxes is based on the classical picture of radiation
with a well-defined classical two-velocity written in the form of light-cone coordinates.
So, one might wonder whether this geometrical approach of decomposition of fluxes
is valid or not. The flux $\langle T^{01} \rangle$ in an inertial frame was given by Eq. \eqref{t01},
and then it
was decomposed into left-right fluxes, which are indeed attributed to the massless property of radiation.
If the Hawking particles were massive, then the chiral decomposition of $\langle T^{01} \rangle$
would be impossible. Moreover, the two-velocity and the normal vector play a role of
transformation matrices from the tortoise coordinate system to the locally inertial coordinate
system as seen from Eq. \eqref{velocity}.
Hence, it is valid to treat these vectors as classical objects
as long as the background metric is fixed in the present semiclassical approximation.
If the back reaction of the geometry were taken into account, then the velocity and the normal vector
should be corrected in accord with quantum corrections.
Secondly, the typical wavelength of outgoing Hawking modes
is comparable to the size of the black hole.
The wavelength of negative ingoing modes
is also the same order of the critical radius $r_c$ at which the two free-fall temperatures become equal.
Now, one might wonder
how much ingoing modes are expected to overlap with the typical
outgoing modes.
This issue will be discussed only in freely falling frames such that
physical quantities are proper ones, for example, the left-right fluxes mean the proper left-right fluxes
as ${\cal F}_{L,R}$ rather than
the flux $\langle T_{\pm\pm} \rangle$ defined in the tortoise coordinates, but
they are related to each other in the way of ${\cal F}_{L,R}=\mp \langle T_{\pm\pm} \rangle /g(r)$ from Eq. \eqref{flux}.
The thermal temperature is usually interpreted as an absolute value of momentum
for a particle. The left temperature from the influx is inversely proportional to the wavelength of ingoing modes.
In particular, the wavelength of ingoing modes is the same as that of outgoing modes at $r_c$.
However,
the wavelength of ingoing modes generically depends on a free-fall position critically, while
the size of the wavelength of outgoing modes is almost comparable to the size of the black hole except for
free-fall positions near the horizon as seen from Fig. \ref{fig:1}.
At the horizon, the influx is divergent at the horizon, and so the left temperature is also
divergent, while the outward flux and the right temperature approach zeroes there.
It means that the wavelength of ingoing modes in the freely falling frame near the horizon
is extremely short in contrast to the very long wavelength of outgoing modes.
On the other hand, at the asymptotic infinity,
the wavelength of ingoing modes approaches infinity while the wavelength of outgoing modes becomes
that of the well-known Hawking modes.
Consequently, the wavelength of ingoing modes is highly sensitive to the free-fall positions
in contrast to that of outgoing modes. As was discussed earlier,
we just speculate that if the divergent ingoing modes
could be regulated properly by taking into account the back reaction of the
geometry, then they would be finite.
Finally, it has been widely believed that a freely falling observer
encounters nothing out of the ordinary when crossing the horizon based on the equivalence principle.
In fact, there is no reason for a freely falling observer to find the divergent proper energy-momentum tensor since
the curvature in the free-fall frame is finite everywhere except the origin.
Thus, the finite proper influx saves
the violation of the equivalence principle except for the horizon where
the firewall is located \cite{Almheiri:2012rt}.



\acknowledgments
I have benefited from discussions with M. Eune, Y. Gim, and E. J. Son, and especially thank W. Israel for introducing his
helpful paper for improvement of the present work.


\bibliographystyle{JHEP}       

\bibliography{references}

\providecommand{\href}[2]{#2}\begingroup\raggedright\begin{thebibliography}{10}

\bibitem{Hawking:1974sw}
S.~W. Hawking, \emph{{Particle Creation by Black Holes}},
  \href{http://dx.doi.org/10.1007/BF02345020}{\emph{Commun. Math. Phys.} {\bf
  43} (1975) 199--220}.

\bibitem{Susskind:1993if}
L.~Susskind, L.~Thorlacius and J.~Uglum, \emph{{The Stretched horizon and black
  hole complementarity}},
  \href{http://dx.doi.org/10.1103/PhysRevD.48.3743}{\emph{Phys. Rev.} {\bf D48}
  (1993) 3743--3761}, [\href{http://arxiv.org/abs/hep-th/9306069}{{\tt
  hep-th/9306069}}].

\bibitem{Almheiri:2012rt}
A.~Almheiri, D.~Marolf, J.~Polchinski and J.~Sully, \emph{{Black Holes:
  Complementarity or Firewalls?}},
  \href{http://dx.doi.org/10.1007/JHEP02(2013)062}{\emph{JHEP} {\bf 02} (2013)
  062}, [\href{http://arxiv.org/abs/1207.3123}{{\tt 1207.3123}}].

\bibitem{Braunstein:2009my}
S.~L. Braunstein, S.~Pirandola and K.~Zyczkowski, \emph{{Better Late than
  Never: Information Retrieval from Black Holes}},
  \href{http://dx.doi.org/10.1103/PhysRevLett.110.101301}{\emph{Phys. Rev.
  Lett.} {\bf 110} (2013) 101301}, [\href{http://arxiv.org/abs/0907.1190}{{\tt
  0907.1190}}].

\bibitem{Page:1993wv}
D.~N. Page, \emph{{Information in black hole radiation}},
  \href{http://dx.doi.org/10.1103/PhysRevLett.71.3743}{\emph{Phys. Rev. Lett.}
  {\bf 71} (1993) 3743--3746}, [\href{http://arxiv.org/abs/hep-th/9306083}{{\tt
  hep-th/9306083}}].

\bibitem{Bousso:2012as}
R.~Bousso, \emph{{Complementarity Is Not Enough}},
  \href{http://dx.doi.org/10.1103/PhysRevD.87.124023}{\emph{Phys. Rev.} {\bf
  D87} (2013) 124023}, [\href{http://arxiv.org/abs/1207.5192}{{\tt
  1207.5192}}].

\bibitem{Nomura:2012sw}
Y.~Nomura, J.~Varela and S.~J. Weinberg, \emph{{Complementarity Endures: No
  Firewall for an Infalling Observer}},
  \href{http://dx.doi.org/10.1007/JHEP03(2013)059}{\emph{JHEP} {\bf 03} (2013)
  059}, [\href{http://arxiv.org/abs/1207.6626}{{\tt 1207.6626}}].

\bibitem{Susskind:2012rm}
L.~Susskind, \emph{{Singularities, Firewalls, and Complementarity}},
  \href{http://arxiv.org/abs/1208.3445}{{\tt 1208.3445}}.

\bibitem{Hossenfelder:2012mr}
S.~Hossenfelder, \emph{{Comment on the black hole firewall}},
  \href{http://arxiv.org/abs/1210.5317}{{\tt 1210.5317}}.

\bibitem{Giddings:2013kcj}
S.~B. Giddings, \emph{{Nonviolent information transfer from black holes: A
  field theory parametrization}},
  \href{http://dx.doi.org/10.1103/PhysRevD.88.024018}{\emph{Phys.Rev.} {\bf
  D88} (2013) 024018}, [\href{http://arxiv.org/abs/1302.2613}{{\tt
  1302.2613}}].

\bibitem{Almheiri:2013hfa}
A.~Almheiri, D.~Marolf, J.~Polchinski, D.~Stanford and J.~Sully, \emph{{An
  Apologia for Firewalls}},
  \href{http://dx.doi.org/10.1007/JHEP09(2013)018}{\emph{JHEP} {\bf 1309}
  (2013) 018}, [\href{http://arxiv.org/abs/1304.6483}{{\tt 1304.6483}}].

\bibitem{Hutchinson:2013kka}
J.~Hutchinson and D.~Stojkovic, \emph{{Icezones instead of firewalls: extended
  entanglement beyond the event horizon and unitary evaporation of a black
  hole}},  \href{http://arxiv.org/abs/1307.5861}{{\tt 1307.5861}}.

\bibitem{Freivogel:2014dca}
B.~Freivogel, \emph{{Energy and Information Near Black Hole Horizons}},
  \href{http://dx.doi.org/10.1088/1475-7516/2014/07/041}{\emph{JCAP} {\bf 1407}
  (2014) 041}, [\href{http://arxiv.org/abs/1401.5340}{{\tt 1401.5340}}].

\bibitem{Unruh:1976db}
W.~G. Unruh, \emph{{Notes on black hole evaporation}},
  \href{http://dx.doi.org/10.1103/PhysRevD.14.870}{\emph{Phys. Rev.} {\bf D14}
  (1976) 870}.

\bibitem{Unruh:1994zw}
W.~G. Unruh, \emph{{Dumb holes and the effects of high frequencies on black
  hole evaporation}},  \href{http://arxiv.org/abs/gr-qc/9409008}{{\tt
  gr-qc/9409008}}.

\bibitem{Casadio:2002dj}
R.~Casadio and L.~Mersini-Houghton, \emph{{Short distance signatures in
  cosmology: Why not in black holes?}},
  \href{http://dx.doi.org/10.1142/S0217751X04016453}{\emph{Int. J. Mod. Phys.}
  {\bf A19} (2004) 1395--1412},
  [\href{http://arxiv.org/abs/hep-th/0208050}{{\tt hep-th/0208050}}].

\bibitem{Israel:2015ava}
W.~Israel, \emph{{Shenanigans at the black hole horizon: pair creation or
  Boulware accretion?}},  \href{http://arxiv.org/abs/1504.02419}{{\tt
  1504.02419}}.

\bibitem{Giddings:2015uzr}
S.~B. Giddings, \emph{{Hawking radiation, the Stefan–Boltzmann law, and
  unitarization}},
  \href{http://dx.doi.org/10.1016/j.physletb.2015.12.076}{\emph{Phys. Lett.}
  {\bf B754} (2016) 39--42}, [\href{http://arxiv.org/abs/1511.08221}{{\tt
  1511.08221}}].

\bibitem{Page:1976df}
D.~N. Page, \emph{{Particle Emission Rates from a Black Hole: Massless
  Particles from an Uncharged, Nonrotating Hole}},
  \href{http://dx.doi.org/10.1103/PhysRevD.13.198}{\emph{Phys. Rev.} {\bf D13}
  (1976) 198--206}.

\bibitem{Giddings:2006sj}
S.~B. Giddings, \emph{{Black hole information, unitarity, and nonlocality}},
  \href{http://dx.doi.org/10.1103/PhysRevD.74.106005}{\emph{Phys. Rev.} {\bf
  D74} (2006) 106005}, [\href{http://arxiv.org/abs/hep-th/0605196}{{\tt
  hep-th/0605196}}].

\bibitem{Tolman:1930zza}
R.~C. Tolman, \emph{{On the Weight of Heat and Thermal Equilibrium in General
  Relativity}}, \href{http://dx.doi.org/10.1103/PhysRev.35.904}{\emph{Phys.
  Rev.} {\bf 35} (1930) 904--924}.

\bibitem{Hartle:1976tp}
J.~B. Hartle and S.~W. Hawking, \emph{{Path Integral Derivation of Black Hole
  Radiance}}, \href{http://dx.doi.org/10.1103/PhysRevD.13.2188}{\emph{Phys.
  Rev.} {\bf D13} (1976) 2188--2203}.

\bibitem{Israel:1976ur}
W.~Israel, \emph{{Thermo field dynamics of black holes}},
  \href{http://dx.doi.org/10.1016/0375-9601(76)90178-X}{\emph{Phys. Lett.} {\bf
  A57} (1976) 107--110}.

\bibitem{Gim:2015era}
Y.~Gim and W.~Kim, \emph{{A Quantal Tolman Temperature}},
  \href{http://dx.doi.org/10.1140/epjc/s10052-015-3765-2}{\emph{Eur. Phys. J.}
  {\bf C75} (2015) 549}, [\href{http://arxiv.org/abs/1508.00312}{{\tt
  1508.00312}}].

\bibitem{Deser:1976yx}
S.~Deser, M.~J. Duff and C.~J. Isham, \emph{{Nonlocal Conformal Anomalies}},
  \href{http://dx.doi.org/10.1016/0550-3213(76)90480-6}{\emph{Nucl. Phys.} {\bf
  B111} (1976) 45}.

\bibitem{Christensen:1977jc}
S.~M. Christensen and S.~A. Fulling, \emph{{Trace Anomalies and the Hawking
  Effect}}, \href{http://dx.doi.org/10.1103/PhysRevD.15.2088}{\emph{Phys. Rev.}
  {\bf D15} (1977) 2088--2104}.

\bibitem{Wald:1999xu}
R.~M. Wald, \emph{{Gravitation, thermodynamics, and quantum theory}},
  \href{http://dx.doi.org/10.1088/0264-9381/16/12A/309}{\emph{Class. Quant.
  Grav.} {\bf 16} (1999) A177--A190},
  [\href{http://arxiv.org/abs/gr-qc/9901033}{{\tt gr-qc/9901033}}].

\bibitem{Eune:2014eka}
M.~Eune, Y.~Gim and W.~Kim, \emph{{Something special at the event horizon}},
  \href{http://dx.doi.org/10.1142/S0217732314502150}{\emph{Mod. Phys. Lett.}
  {\bf A29} (2014) 1450215}, [\href{http://arxiv.org/abs/1401.3501}{{\tt
  1401.3501}}].

\bibitem{Callan:1992rs}
C.~G. Callan, Jr., S.~B. Giddings, J.~A. Harvey and A.~Strominger,
  \emph{{Evanescent black holes}},
  \href{http://dx.doi.org/10.1103/PhysRevD.45.R1005}{\emph{Phys. Rev.} {\bf
  D45} (1992) 1005--1009}, [\href{http://arxiv.org/abs/hep-th/9111056}{{\tt
  hep-th/9111056}}].

\bibitem{Boulware:1974dm}
D.~G. Boulware, \emph{{Quantum Field Theory in Schwarzschild and Rindler
  Spaces}}, \href{http://dx.doi.org/10.1103/PhysRevD.11.1404}{\emph{Phys. Rev.}
  {\bf D11} (1975) 1404}.

\bibitem{BoschiFilho:1991xz}
H.~Boschi-Filho and C.~P. Natividade, \emph{{Anomalies in curved space-time at
  finite temperature}},
  \href{http://dx.doi.org/10.1103/PhysRevD.46.5458}{\emph{Phys. Rev.} {\bf D46}
  (1992) 5458--5466}.

\bibitem{Visser:1996ix}
M.~Visser, \emph{{Gravitational vacuum polarization. 3: Energy conditions in
  the (1+1) Schwarzschild space-time}},
  \href{http://dx.doi.org/10.1103/PhysRevD.54.5123}{\emph{Phys. Rev.} {\bf D54}
  (1996) 5123--5128}, [\href{http://arxiv.org/abs/gr-qc/9604009}{{\tt
  gr-qc/9604009}}].

\bibitem{Singleton:2011vh}
D.~Singleton and S.~Wilburn, \emph{{Hawking radiation, Unruh radiation and the
  equivalence principle}},
  \href{http://dx.doi.org/10.1103/PhysRevLett.107.081102}{\emph{Phys. Rev.
  Lett.} {\bf 107} (2011) 081102}, [\href{http://arxiv.org/abs/1102.5564}{{\tt
  1102.5564}}].

\bibitem{Hawking:2014tga}
S.~W. Hawking, \emph{{Information Preservation and Weather Forecasting for
  Black Holes}},  \href{http://arxiv.org/abs/1401.5761}{{\tt 1401.5761}}.

\bibitem{Visser:2014zqa}
M.~Visser, \emph{{Physical observability of horizons}},
  \href{http://dx.doi.org/10.1103/PhysRevD.90.127502}{\emph{Phys. Rev.} {\bf
  D90} (2014) 127502}, [\href{http://arxiv.org/abs/1407.7295}{{\tt
  1407.7295}}].

\bibitem{Mersini-Houghton:2014zka}
L.~Mersini-Houghton, \emph{{Backreaction of Hawking Radiation on a
  Gravitationally Collapsing Star I: Black Holes?}},
  \href{http://dx.doi.org/10.1016/j.physletb.2014.09.018}{\emph{Phys. Lett.}
  {\bf B738} (2014) 61--67}, [\href{http://arxiv.org/abs/1406.1525}{{\tt
  1406.1525}}].

\bibitem{Mersini-Houghton:2014cta}
L.~Mersini-Houghton and H.~P. Pfeiffer, \emph{{Back-reaction of the Hawking
  radiation flux on a gravitationally collapsing star II}},
  \href{http://arxiv.org/abs/1409.1837}{{\tt 1409.1837}}.

\bibitem{Nomura:2016qum}
Y.~Nomura and N.~Salzetta, \emph{{Why Firewalls Need Not Exist}},
  \href{http://arxiv.org/abs/1602.07673}{{\tt 1602.07673}}.

\end{thebibliography}\endgroup

\end{document}